# Pulse Parameter Optimization Method for Ultra High Dose Rate Electron Treatment

S. Jain[1], A. Cetnar[1], J. Woollard[1], N. Gupta[1], D. Blakaj[1], A. Chakravarti[1]*, and A. S. Ayan[1]*

[1]The Department of Radiation Oncology, the James, The Ohio State University
*A. Chakravarti and A. S. Ayan are joint senior authors as the Clinical and Physics Leads respectively.

## Abstract

Purpose: Commercial UHDR platforms deliver Ultra-High Dose Rate (UHDR) doses at discrete combinations of pulse parameters including pulse width (PW), pulse repetition frequency (PRF) and number of pulses (N), which dictate unique combinations of dose and dose rates. Currently, obtaining pulse parameters for the desired dose and dose rate is a cumbersome manual process involving creating, updating, and looking up values in large spreadsheets for every treatment configuration. The purpose of this work is to present a pulse parameter optimizer application to match intended dose and dose rate precisely and efficiently.

Methods: Dose and dose rate calculation have been described for a commercial electron FLASH platform. A constrained optimization for the dose and dose rate cost function was modelled as a mixed integer problem in MATLAB (The MathWorks Inc., Version9.13.0 R2022b, Natick, Massachusetts). The beam and machine data required for the application were acquired using GafChromic film and Alternating Current Current Transformers (ACCTs). Variables for optimization included dose per pulse (DPP) for every collimator at a specific treatment configuration, PW and PRF measured using ACCT, and airgap factors.

Results: Using PW, PRF, N and airgap factors as the parameters, the application was created to optimize for dose and dose rate. Largely automating dose and dose rate calculation reduces safety concerns associated with manual look up and calculation of these parameters, especially when many subjects at different doses and dose rates are to be safely managed.

Conclusion: A pulse parameter optimization application was built in MATLAB for a commercial electron UHDR platform to increase efficiency in the dose, dose rate, and pulse parameter prescription process.

## 1. Introduction

Ultra-high dose rate radiotherapy delivers treatment in millisecond timescales. Multiple pre-clinical models including mice, zebrafish, cats, and pigs [1] [2] [3] [4] , have shown decreased side effects on healthy tissues, when compared to conventional dose rates delivered in minutes. Tumor control studies with various cell lines have shown equivalent tumor control with conventional and UHDR irradiation [5] [6] [7] [8]. This biological phenomenon of reducing normal tissue toxicity while maintaining isoeffective tumor control has been called the FLASH effect.

From a treatment planning and prescription perspective, unlike conventional radiotherapy specifying dose, FLASH RT considers dose rate as an additional variable. The FLASH effect has been largely seen at > 10 Gy total dose and > 40 Gy/s dose rates [9] [10]. However, this phenomenon is not binary, as varying magnitudes of the FLASH effect have been reported with increasing dose rate [11], and it may also be dependent on beam-specific pulse parameters [12]. As more commercial platforms for UHDR production emerge for exploration of FLASH RT [13-16], it is important for pre-clinical studies to reproduce the dose and dose rate combinations that have produced the FLASH effect and further explore in this space to increase magnitude of the effect.

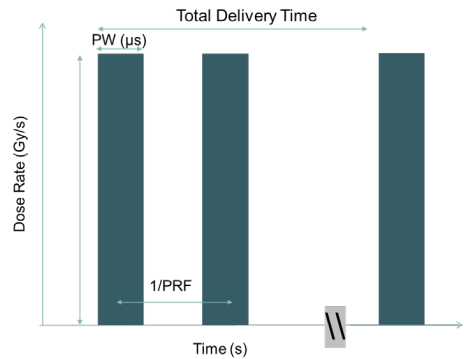

Figure 1: Adapted from Wilson et al, Front. Oncol., 2020 [9]. Ideal pulsed FLASH beam, where the width of an individual pulse is the pulse width, and the frequency of pulses is the pulse repetition frequency.

Most pulsed electron UHDR platforms allow for manual settings of pulse parameters such as pulse width (PW), pulse repetition frequency (PRF), grid tension, number of pulses (N), etc. (Figure 1). Attaining specific combinations of dose and dose rate is not trivial as these two factors become interdependent when the electron pulse structure is modified. The Mobetron (IntraOp, Sunnyvale, CA, USA) has been used for pre-clinical FLASH studies at our institution. UHDR delivery requires manual settings of PW, PRF, and N available through discrete settings on dials, which manipulate the dose per pulse (DPP), dose rate, and total dose. Changing the source to surface distance (SSD) will also change the dose per pulse, total dose, and dose rate. Since the settings available for these parameters are limited and discrete, one can only achieve certain combinations of dose and dose rates. This creates an optimization problem to match the intended dose and dose rate for treatment with what is achievable through the machine variables for a particular treatment configuration. Handling the creation and updates of numerous long spreadsheets with all these parameter combinations is inefficient and prone to human error. In this work, we present a treatment planning tool to optimize these parameters to achieve the intended dose and dose rate available using the Mobetron, which can be extended to variables available for other UHDR pulsed electron accelerators.

2. Methods

Dosimetry is a challenge in the UHDR realm due to increased ion recombination in standard dosimeters such as ion chambers [17]. Film is the standard dosimeter for UHDR beams [18] due to its absence of dose per pulse dependency [18]. Since UHDR mode in the FLASH Mobetron requires manual settings of pulse parameters, it is important to commission and monitor these parameters on pulse waveforms. ACCTs (Alternating Current Current Transformers) in the head of the Mobetron have shown promise in providing pulse parameter information [19-21] as well as for real time dosimetry. This work utilizes EBT-XD GafChromic film (Ashland Advanced Material, NJ, USA) for dose measurements and ACCT measurements for characterizing and reporting of pulse parameters. The following sections describe dose and dose rate calculation based on machine parameters and the beam data required to run the optimization tool.

a. Description of Treatment Configurations and Pulse Parameters

The FLASH Mobetron in the IORT configuration has two available SSD configurations for irradiation using collimators made from Delrin® (polyoxymethylene) material ranging from 2.5 cm to 6 cm in diameter. Figure 2 shows the SSD configurations available: configuration A provides an SSD of 18.3 cm through direct mounting of the collimator on the exit window of the gantry and configuration B is achieved through an applicator and provides an SSD of 35 cm. SSD configurations have impact not only on the dose per pulse and dose rate, but also on the shape of profiles and Percent Depth Dose (PDD). This work includes characterization and beam parameter optimization for the SSD = 18.3 cm configuration, to provide a framework that can be replicated for other configurations. Beam data was acquired for every collimator at this configuration using EBT-XD GafChromic film.

The Mobetron allows pulse control by manual settings of pulse repetition frequency (PRF) between 10 and 60 Hz and pulse widths (PW) between 1 and 4 µs through dials shown in Figure 2. There are also two ACCTs (Bergoz Instrumentation, France) in the head of the machine, one below the primary scattering foil and the other below the secondary scattering foil (Figure 2), which produce signal via electromagnetic induction as the electron beam passes through them. We use a digital oscilloscope (Picoscope 5000, PicoTechnology, UK) to record pulse waveforms and analyze in MATLAB to obtain measured PWs (FWHM of pulses) and area under the pulses. The latter can be correlated to charge by applying voltage to current conversion coefficients supplied by the vendor.

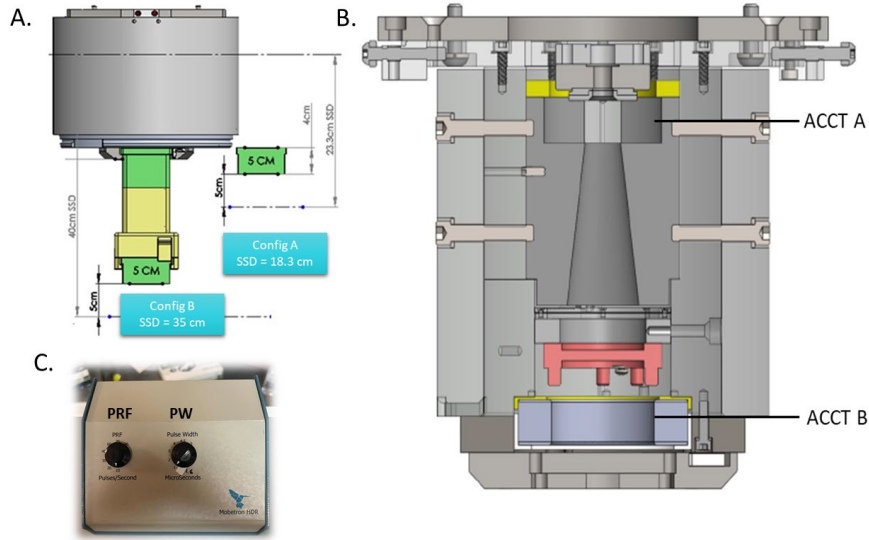

Figure 2: A. SSD Configurations on the Mobetron. Config A is achieved by directly mounting the collimator on the exit window, and Config B is achieved through an additional applicator. B. Mobetron gantry with two ACCTs added for beam current sensing. Drawings received from IntraOp (Sunnyvale, CA, USA). C. PRF and PW dials to allow pulse parameter manipulation for UHDR beams.

b. Dose and dose rate calculation

Dose calculation for the FLASH Mobetron is different from standard MU based dose calculations defined in AAPM TG71 [22]. The dose is given by,

$$D\ (Gy) = N * DPP\ (60\ Hz, 4\ \mu s) * PWF * g\ (cone, SSD) \quad (1)$$

where $D$ is the total dose, $N$ is the number of pulses, $DPP$ is the dose per pulse measured at 60 Hz PRF and 4 µs PW setting for the cone being used at SSD = 18.3 cm, $PWF$ is a pulse width factor to account for relative differences in DPP between pulse width settings, and $g$ is an airgap dependent factor unique to a particular collimator and configuration. Dose rate is then calculated using,

$$\dot{D}\left(\frac{Gy}{s}\right) = \frac{D\ (Gy)}{\frac{N-1}{PRF\ (Hz)} + PW\ (\mu s)} \quad (2)$$

Dose per pulse was measured using GafChromic EBT-XD film for each collimator and configurations (SSD 18.3 and 35 cm) at the depth of $d_{max}$. The physical pulse widths were measured by the ACCTs, which are different from the nominal settings on the PW dials. It should be noted that these values are specific to every machine. The measured values at 50% level of the ACCT pulses are used in calculations. Pulse width factors are obtained through film measurements in the same setup, while only varying the nominal PW setting.

$$PWF = \frac{Dose\ (PW, PRF)}{Dose\ (4\ \mu s, 60\ Hz)} \quad (3)$$

To characterize dose distributions at airgaps, multiple film measurements parallel to the irradiation beam per collimator were obtained at 1-2 cm increment airgaps up to 10 cm. The parallel film irradiations were performed using an in-house 3D printed vertical film holder that allowed placement of the film level with the surface of the water [23]. A laser distance meter was used to measure the set airgap between the film/water surface and the collimator. In this way, we were able to obtain dose per pulse, PDDs, and profiles for various airgaps with one film measurement per airgap. The doses extracted at $d_{max}$ were normalized to g = 0 measurement. These were then plotted against airgap and fitted with a second order polynomial function with the intercept set to 1 (since at zero airgap, this factor would be 1). For field sizes < 4cm, airgaps < 3 cm were excluded due to considerably decreased therapeutic depth (R80) and higher hotspots in the profiles as seen in figure 2(b) in the supplementary materials. Data containing profiles and PDDs at various airgaps for the largest and smallest collimator sizes can also be found in the supplementary materials.

c. Parameter Optimization

The FLASH Mobetron delivers the requested number of pulses with the selected pulse repetition frequency and the pulse width. No partial pulse delivery is achievable. These parameters, along with the cone size and the irradiation geometry determine the dose and dose rates. While different machines have slightly varying configurations, our institution's Mobetron can be programmed to run with PRFs of 10, 20, 30, 45, and 60 Hz and nominal PWs of 1, 1.6, 2, 3 and 4 microseconds.

The objective function is set up to best achieve the desired dose and the dose rates within the parameter space mentioned above as,

$$min\left\{\left(\frac{D_{achived}}{D_{requested}} - 1\right)^2 + \left(\frac{\dot{D}_{achived}}{\dot{D}_{requested}} - 1\right)^2\right\}, \qquad (4)$$

where $D$ and $\dot{D}$ are given by equations (1) and (2), subject to
$PRF \in \{10,20,30,45,60\}\,Hz, \; PW \in \{1,1.6,2,3,4\}\,\mu s, \; N \in \mathbb{Z}, \; g \in \{1 - 100\}\,mm$

The constrained optimization for the objective function was setup as a mixed integer problem in MATLAB since only integer number of pulses could be delivered, and a discrete value of PRF and PWs could be selected by the optimizer while the airgap distance could be set to a value between 0 and 10 cm.

3. Results
  3.1 Beam and pulse data

Table 1 shows the DPP at $d_{max}$ at PRF = 60 Hz and nominal PW = 4 μs obtained with GafChromic film for different cones. These were acquired at the depth of $d_{max}$ and airgap of 0 cm for cone sizes > 4 cm, and an airgap of 3 cm otherwise due to unfavorable beam profiles at shorter SSDs and smaller cone sizes (data included in supplementary materials). Three films were irradiated at every setting, and doses extracted were averaged. Table 2 lists the measured pulse widths obtained from ACCT data at the full width at half maximum of a pulse averaged across PRFs. Film doses were acquired at all combinations of PW and PRFs and were normalized to the 60 Hz and 4 μs setting and averaged across PRFs to obtain PW factors listed in Table 2.

Table 1: Dose per pulse for circular collimators at SSD Configurations A. PRF = 60 Hz, PW = 4 us. Errors correspond to the standard deviation over three measurements.

| Collimator (cm) | SSD = 18.3 cm (cGy) | Airgap (cm) |
|---|---|---|
| 2.5 | 771 ± 7.72 | 3 |
| 3 | 770 ± 6.28 | 3 |
| 4 | 735 ± 2.40 | 3 |
| 5 | 846 ± 8.62 | 0 |
| 6 | 819 ± 4.74 | 0 |

Table 2: Measured pulse widths (ACCTs) and relative pulse width factor (obtained from film measurements). Errors correspond to standard deviation over measurements across all PRFs.

| Nominal Pulse Width | Measured Pulse Width (FWHM, µs) | Pulse Width Factors |
|---|---|---|
| 1 | 1.11 ± 0.02 | 0.40 ± 0.01 |
| 1.6 | 1.72 ± 0.02 | 0.56 ± 0.01 |
| 2 | 2.13 ± 0.02 | 0.68 ± 0.01 |
| 3 | 3.09 ± 0.01 | 0.93 ± 0.02 |
| 4 | 3.44 ± 0.00 | 1.00 ± 0.02 |

Figure 3 demonstrates the space of achievable combinations of dose and dose rate as a function of number of pulses and PW at a set PRF of 60 Hz and dose per pulse corresponding to a 6 cm cone at Configuration A. Maximum dose rate that can be achieved in a single pulse delivery is of the order of $10^6$ Gy/s, which then drops to $10^2$ Gy/s mean dose rates as the number of pulses increases because PRF starts dominating the dose rate calculation.

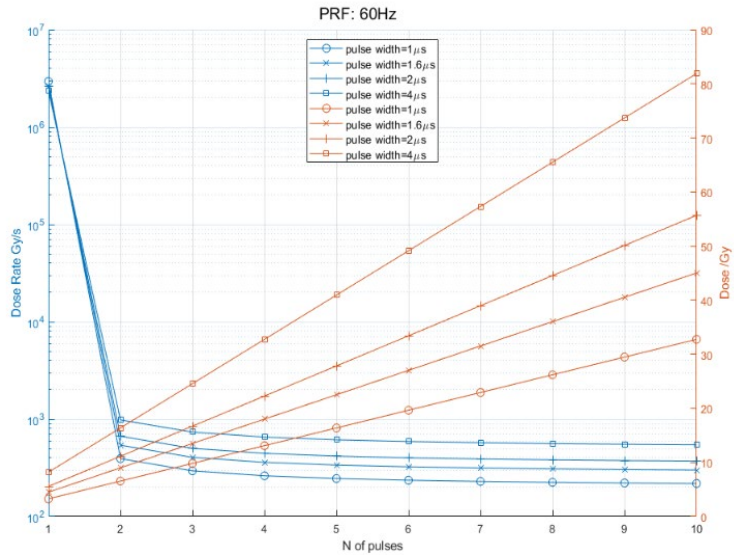

*Figure 3: Dose (right y axis) and dose rate (left y axis) as a function of number of pulses (x axis) for different pulse widths at a set PRF of 60 Hz. Dose per pulse for a 6cm collimator at SSD Configuration A was used for calculations. Markers represent the discrete values of dose and dose rates that can be achieved.*

### 3.2 Airgap as a Variable

Film measurements parallel to the beam direction were acquired for every collimator at various airgaps between 0 and 10 cm (0, 0.5, 1, 2, 3, 4, 6, 8, 10 cm) using the vertical film holder apparatus described in the methods. The doses extracted at $d_{max}$ normalized to either g = 0 or g = 3 cm were plotted against airgap and fitted with a second order polynomial, as shown in figure 4 for cone sizes of 2.5 cm and 6 cm.

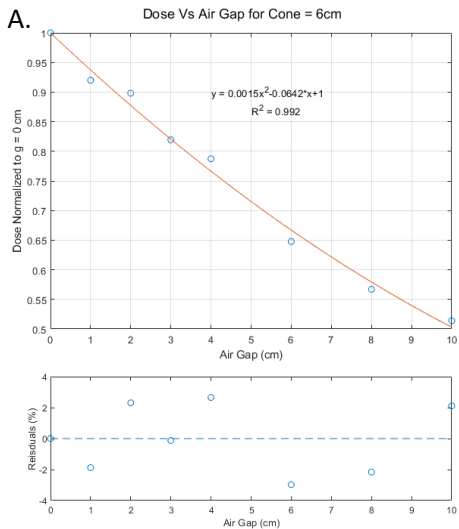
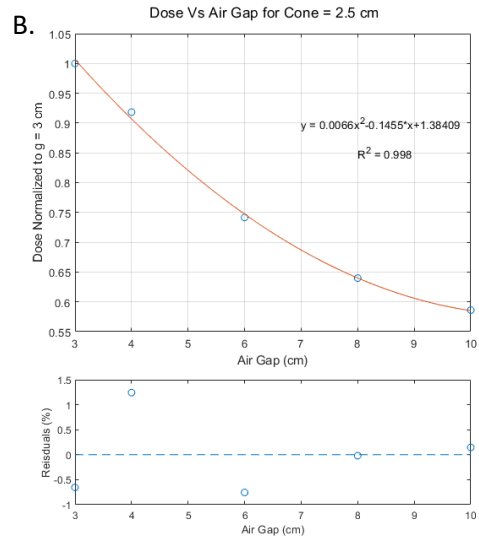

*Figure 4: Dose fall off as a function of airgap between the collimator and phantom surface. (a) shows the fall of for 6 cm diameter cone normalized to the g = 0 maximum dose per pulse along with the corresponding second order polynomial fit and residuals. (b) shows the dose fall off for 2.5 cm diameter cone. Here, the normalizing condition was set at a 3 cm air gap due to unfavorable beam profiles and PDDs at shorter air gaps. Residuals show the percent differences between the data points and the fits.*

### 3.3 Pulse parameter optimizer

A MATLAB-based GUI was built for the optimizer. The application takes desired dose, dose rate, and cone size as input, and returns number of pulses, pulse width, PRF, and airgap needed to achieve the closest matching dose and dose rate values. It also returns the dose and dose rates achieved through the optimized parameters. Table 3 summarizes the data needed for the optimizer.

*Table 3: Input data needed for Pulse Parameter Optimization*

| Beam/Pulse Data | Description |
| --- | --- |
| Dose per pulse | Measured using GafChromic film for every treatment configuration |
| Pulse Widths | Pulse Widths measured using ACCT waveforms |
| Pulse Width Factors | Relative doses with change in pulse width obtained using film |
| Dose fall-off with airgap | Dose per pulse fall off with increasing airgap and corresponding beam data to assess changes in PDDs and Profiles |

The optimizer converged to results in less than a minute. It was tested with known dose and dose rate combinations from previously used spreadsheets based on manual calculations and was able to reproduce the same pulse parameters as calculations. Quality assurance of the optimizer with known combinations is recommended prior to each use. This MATLAB code is available upon request, but data used to generate the optimization is machine specific.

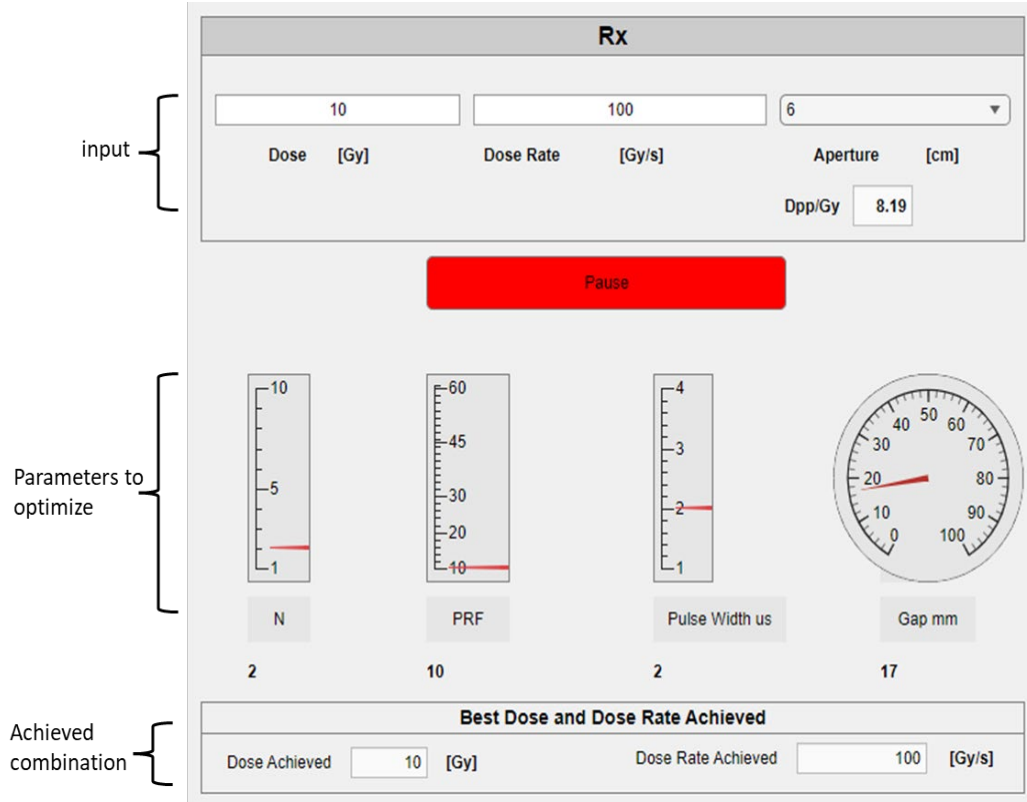

*Figure 5: Screen capture of the Pulse Parameter Optimizer GUI. This example shows a desired dose and dose rate of 10 Gy and 100 Gy/s respectively, with a cone size of 6 cm. The optimization returns 2 pulses, 10 Hz for PRF, 2 µs Nominal PW and 17 mm airgap to achieve the prescribed dose and dose rate.*

4. Discussion

Exploration of the dose and dose rate space for the optimal combination of the FLASH effect is essential for pre-clinical studies and can guide standardization for future clinical use. Since dose and dose rate are not independent variables for the FLASH Mobetron, there is a need to understand what combinations are achievable with all pulse parameters available for guiding the decision-making process for researchers and clinicians. The pulse parameter optimization tool helps navigate multiple variables to achieve the desired dose and dose rates more precisely and efficiently. Since the Mobetron provides upwards of 8 Gy per pulse, deviations in pulse parameter calculation through manual look up could lead to large differences in dose, which is a serious safety concern. Automating this task greatly reduces human error in maintaining spreadsheets and look-up tables for various combinations of pulse parameters for every collimator.

The current work is developed for Delrin collimators at SSD = 18.3 cm and can be adapted to other geometries with additional beam data. The current version of the optimizer also gives equal weight to all the parameters available, but priorities can be assigned to favor one parameter over another. Examples could be if the dose variable was prioritized over dose rate or if it was desired to fix the pulse width while varying other parameters.

The use of distance as an additional variable has also been explored to achieve more granularity in dose. The effective SSD and gap factor methods as described in AAPM TG-71 [22] are designed for a conventional linac at a much larger SSD (= 100 cm). They use a single $d_{max}$ value for PDDs across all airgaps for calculation of the effective SSD or gap factors. Since the depth of $d_{max}$ changed considerably with airgap for smaller fields, collimator specific fits of dose fall off with airgap were obtained instead to avoid the uncertainty in effective SSD calculations. Irradiating film parallel to the beam direction at various airgaps allowed for assessment of changes in beam characteristics, based on which smaller airgaps for < 4 cm cone sizes were excluded due to higher hotspots and reduced therapeutic depth. Since the Mobetron does not have an internal optical distance indicator display or couch positions, an external laser device pointing towards a fixed spot on the gantry was used to set the airgaps. The airgap data is also limited by the uncertainty associated with the laser pointer (± 1mm).

It should be noted that different versions of the Mobetron have slightly different configurations for PW, PRF, and collimation. Based on these parameters, there are also differences in outputs across machines. Therefore, every user must characterize the parameters mentioned in Table 3 for use with the software. The current dose calculation formalism has limitations since it does not account for potential changes in beam energy with pulse width, where beam energy could decrease with increasing pulse width due to the beam loading effect. The current version of the optimizer also does not account for pulse-to-pulse variations in output, which can amount to up to 6% between the first and subsequent pulses, as reported for our machine [24]. Users must also characterize any changes in dose per pulse with PRF, since lower doses and lower stability at higher PRFs have been observed. Since every machine is characterized differently, these variations must be studied and characterized by every user. Film dosimetry adds additional uncertainty to all the data, which has been reported up to 4% for EBT3 films between 3 and 17 Gy [18]. With the development of new UHDR-compatible active detectors for beam data collection, this uncertainty may be greatly reduced.

5. Conclusion

A MATLAB-based pulse parameter optimization software for a commercial platform producing pulsed UHDR electron beams was developed. This tool helps navigate the dose and dose rate space to obtain the best pulse parameters required to achieve the intended dose and dose rate combination. By automating this process, users can more efficiently and precisely match intended dose and dose rate, and greatly reduce human error in manual look up of pulse parameters for calculations.